\documentclass[pra,twocolumn,,draft,showpacs,tightenlines]{revtex4}
\usepackage{amssymb}
\usepackage{amsmath}
\usepackage{colordvi}

\begin{document}

\title{State estimation from pair of conjugate qudits}
\author{Xiang-Fa Zhou}
\email{xfzhou@mail.ustc.edu.cn}
\author{Yong-Sheng Zhang}
\email{yshzhang@ustc.edu.cn}
\author{Guang-Can Guo}
\email{gcguo@ustc.edu.cn}

 \affiliation{\textit{Key Laboratory of
Quantum Information, University of Science and Technology of
China, Hefei, Anhui 230026, People's Republic of China}}

\begin{abstract}
We show that, for $N$ parallel input states, an antilinear map
with respect to a specific basis is essentially a classical
operation. We also consider the information contained in
phase-conjugate pair $|\phi \rangle |\phi^* \rangle$, and prove
that there is more information about a quantum state encoded in
phase-conjugate pair than in parallel pair.
\end{abstract}

\maketitle

Wigner's theorem says that symmetry transformations in quantum
mechanics must be unitary or antiunitary \cite{wigner}. The
identity is unitary and symmetry,  hence any symmetry that can
become the identity by continuously changing a parameter has a
unitary representation. On the contrary antiunitary maps are not
connected to the identity, so it is not suitable to describe
quantum physics. Therefore nature chooses unitary dynamics as its
reasonable description. Nevertheless antiunitary transformations
still have many interesting properties. The well-known Kramer's
degeneracy comes from the time-reverse symmetry of quantum systems
which contain an odd number of fermions \cite{quantumbook}. Such
time-reversal transformation is antiunitary. Usually an
antiunitary operator can be decomposed into an antilinear
transformation multiplied by a unitary operator. Many strange
properties come from the antilinear part of such an operator.
Recent progress in quantum information reveals that antilinear
operator may play an important role during the study of quantum
entanglement. In fact the famous Positive Partial Transpose (PPT)
criterion \cite{ppt}, where antilinear map acts on the second
particle of a bipartite density operator, provides a useful
condition for testing quantum separability. Also antilinear map
can be used as a useful technique to construct superoperators
\cite{preskill}. In the case of 2-dimensional Hilbert space,
antilinear map is directly related to the universal-flip of a
quantum state \cite{buzek}. For high dimensional case,
universal-flip operator doesn't exist. However, the map $|\phi
\rangle=\sum_i \alpha_i |i\rangle \rightarrow |\phi^*
\rangle=\sum_i \alpha_i^* |i\rangle$ with respect to a specific
basis $\{ |i\rangle \}$ still has many interesting properties.
Here the state vector $|\phi^* \rangle$ is often called as the
phase-conjugate state of $|\phi \rangle$.


In the case of continuous quantum variables, Cerf \emph{et al.}
\cite{cerf} have pointed out that phase conjugate of an unknown
Gaussian state can be realized by measurement procedure. We think
that a similar result exists for any finite dimension case. In
this paper, we simply prove that an antilinear map is essentially
a classical operation. That is, if the input states are composed
of $N$ copies of $|\phi\rangle$, a quantum operation can be
implemented by classical ways. We next consider the information of
a quantum state $|\phi \rangle$ contained in phase-conjugate pair
$|\phi \rangle |\phi^* \rangle$. We find that there is more
information about a quantum state encoded in phase-conjugate pair
than in parallel pair. In the two-level case, when the number of
the output copies is sufficiently large, quantum cloning with two
antiparallel spins $|\vec{n},-\vec{n}\rangle$ can get higher
fidelity than with parallel spins \cite{anti-cloning}. Our proof
reveals that such result still holds in the high-dimension case.

Consider a $d$-level system with $N$ copies. The whole state of
such system can be expressed as $|\phi \rangle^{\otimes N}$ and
$|\phi \rangle \in {\cal H}$ (Here and the following, without loss
of generality, we assume that the Hilbert space ${\cal H}$ is
$C^d$). The space spanned by all these states, which is often
called as the ``Bose subspace'' of ${\cal H}^{\otimes N}$ and
denoted by ${\cal H}_+^{\otimes N}$ \cite{werner}, is invariant
under permutation $S_N$. Our aim is just to find a
trace-preserving completely positive (CP) map $\xi: {\cal
H}_+^{\otimes N}\rightarrow{\cal H}$, which can maximize the mean
fidelity
\begin{eqnarray}
F=\int d\phi F(\phi)=\int d\phi \mbox{Tr}(|\phi^*\rangle \langle
\phi^*| \xi(|\phi\rangle \langle \phi|^{\otimes N})).
\label{fidelity1}
\end{eqnarray}

For a quantum operation $\xi$ it is always possible to find a set
of operators which satisfy $\xi(\rho)=\sum_\mu A_\mu \rho
A_\mu^\dag$ with the normalization condition $\sum_\mu A_\mu^\dag
A_\mu =I$. This is also known as Kraus representation \cite{kraus}
of quantum operation. By substituting this into Eq.
(\ref{fidelity1}) we can obtain
\begin{eqnarray}
F&=& \int d\phi \sum_\mu \mbox{Tr} [ A_\mu |\phi\rangle\langle
\phi| ^{\otimes N} A_\mu^\dag |\phi^*\rangle \langle \phi^*|] .
\label{fidelity11}
\end{eqnarray}
Before proceeding let us introduce the natural isomorphism between
operators $A:{\cal H}_1 \rightarrow {\cal H}_2$ and vectors $|A
\rangle\rangle$ in ${\cal H}_2 \otimes {\cal H}_1$ which is
defined by
$|A\rangle\rangle=\sum_{i,j}A_{ij}|i\rangle_2|j\rangle_1$ (Here
${\cal H}_1$ and ${\cal H}_2$ are not required to have the same
dimension). This method has been used in many related works
\cite{ariano,buscemi} and can greatly simplify the question we
consider here. It is not difficult to testify that the following
identities are satisfied
\begin{eqnarray}
&&M \otimes N|A\rangle\rangle =|M A N^\tau\rangle\rangle,\\
&&\mbox{Tr}_1[|A\rangle\rangle\langle\langle B|]=A B^\dag,\\
&&\mbox{Tr}_2[|A\rangle\rangle\langle\langle B|]= A^\tau B^*, \\
&&\mbox{Tr}[A M_1 A^\dag M_2]=
\mbox{Tr}[|A\rangle\rangle\langle\langle A| M_2 \otimes M_1^\tau].
\end{eqnarray}
Here $\tau$ and $*$ represent the transposition and complex
conjugation with respect to the fixed basis, while $\mbox{Tr}_i$
denotes the partial trace over the Hilbert space ${\cal H}_i$.
Hence by introducing this new notation Eq. ($\ref{fidelity11}$)
now can be rewritten as
\begin{eqnarray} \label{f4}
F &=& \sum _\mu \mbox{Tr}[ |A_\mu^* \rangle\rangle\langle\langle
A_\mu^*| \int d\phi |\phi\rangle\langle\phi|^{\otimes N+1}]
\nonumber \\
&=& \frac{1}{d[N+1]}\sum _\mu \mbox{Tr}[ |A_\mu^*
\rangle\rangle\langle\langle A_\mu^*| I_{{\cal H}_+^{\otimes
N+1}}],
\end{eqnarray}
where $d[N+1]=\frac{(N+d)!}{(d-1)!(N+1)!}$ represents the
dimension of the ``Bose subspace'' ${\cal H}_+^{\otimes N+1}$.
Since $\{A_\mu\}$ composes a complete quantum operation, under the
natural isomorphism the normalization condition becomes
\begin{eqnarray} \label{normal}
\sum_{\mu} \mbox{Tr}_{\cal
H}(|A_{\mu}^\dag\rangle\rangle\langle\langle
A_{\mu}^\dag|)=I_{{\cal H}_+^{\otimes N}}
\end{eqnarray}
Substituting Eq. ($\ref{normal}$) into Eq. ($\ref{f4}$) one can
easily find that
\begin{eqnarray}
F\leq \frac{1}{d[N+1]} \mbox{Tr}[
|A_{\mu}^\dag\rangle\rangle\langle\langle A_{\mu}^\dag| ] =
\frac{d[N]}{d[N+1]}=\frac{N+1}{N+d}. \label{fidelity}
\end{eqnarray}

Interestingly the right-hand-side of Eq. ($\ref{fidelity}$) is
just the optimal fidelity for state estimation from $N$ parallel
input copies \cite{brub}. Usually quantum physics is governed by
unitary operations. Any physical accessible operations can be
understood from the unitary evolution plus projective measurements
process. In the most cases, quantum operations have been
demonstrated superior to their classical correspondence. However
Eq. ($\ref{fidelity}$) reveals that the fidelity of antilinear
operation is bounded by the amount of classical information
distillable from the input states. This means one can construct
the phase-conjugate states of the inputs through a classical
measurement-based scenario. It should be addressed that the
irreducibility of the input state space plays an important role in
the derivation (see Eq. ($\ref{f4}$)). Recently it has been
pointed out by Buscemi \emph{et al} \cite{buscemi1} that for
equatorial states the optimal phase covariant time-reversal states
cannot be achieved via a measurement-preparation procedure.

In the case of 2-level system, Positive Operator Value Measure
(POVM) acting on $|\phi\rangle|\phi^*\rangle$ can get more
information than two parallel states $|\phi\rangle|\phi\rangle$
\cite{popscu,massar}. This result is often considered as an
evidence of ``nonlocality without entanglement''. In fact, as we
will mention below, in high dimensional system $d\geq3$, such
result will still hold. The method we use here can be regarded as
a generalized version of Ref. \cite{popscu,massar} in
high-dimension case.

For phase-conjugate pair $|\phi\rangle|\phi^*\rangle$, the density
matrix is connected with $\rho(\phi,\phi)=|\phi\rangle\langle
\phi|^{\otimes 2}$ by
\begin{eqnarray}
\rho(\phi,\phi^*)=|\phi\rangle \langle\phi| \otimes
|\phi^*\rangle\langle\phi^*|=\rho(\phi,\phi)^{\mbox{\tiny
$\sim$}\mbox{\tiny$T$}},
\end{eqnarray}
where $^{\mbox{\tiny $\sim$}\mbox{\tiny $T$}}$ denotes the partial
transpose of the second particles. Suppose there exists a set of
Hermitian operators $\hat{a}_i$ which satisfies the following
identity $\sum_i \hat{a}_i=I$. Since $\hat{a}_i$ is Hermitian ,
one can always express it as
\begin{eqnarray}
\hat{a}_i&=&w^{(i)} I \otimes I + \sum_{m,n}t_{mn}^{(i)}
\hat{\lambda}_m\otimes \hat{\lambda}_n \nonumber \\
&\mbox{ }& +\sum_m ( r_m^{(i)} \cdot \hat{\lambda}_m \otimes I+
s_m^{(i)} \cdot I \otimes \hat{\lambda}_m),
\end{eqnarray}
where $\hat{\lambda}_m$ represent the generators of the unitary
group SU($d$). The explicit form of $\hat{\lambda}_m$ can be found
elsewhere \cite{book}. When $\hat{a}_i\geq 0$, which indicates
that $\hat{a}_i$ are physical accessible operations, the set
$\{\hat{a}_i\}$ constitutes a complete POVM and
$\mbox{Tr}(\rho(\phi,\phi)\hat{a}_i))$ corresponds to the
probability of getting the measurement outcome $i$. Now consider
the passive transformation of $^{\mbox{\tiny $\sim$}\mbox{\tiny
$T$}}$ on $\{\hat{a}_i\}$, which is defined by
\begin{eqnarray}
\mbox{Tr}(\hat{a}\rho^{\mbox{\tiny $\sim$}\mbox{\tiny
$T$}})=\mbox{Tr}(\hat{a}^{\mbox{\tiny $\sim$}\mbox{\tiny $T$}} \!
\rho).
\end{eqnarray}
If $\hat{a}_i^{\mbox{\tiny $\sim$}\mbox{\tiny $T$}} \geq 0$ for
all $i$, $\{ \hat{a}_i^{\mbox{\tiny $\sim$}\mbox{\tiny $T$}} \}$
constitute a complete POVM for the input state
$\rho(\phi,\phi^*)$. The probability of getting the outcome $i$
for input state $|\phi \rangle |\phi^* \rangle$ now becomes
$\mbox{Tr}\bigl(\hat{a}_i^{\mbox{\tiny $\sim$}\mbox{\tiny $T$}}
\rho(\phi,\phi^*)\bigr)=\mbox{Tr} \bigl( \hat{a}_i \rho(\phi,\phi)
\bigr)$. Therefore by introducing the $^{\mbox{\tiny
$\sim$}\mbox{\tiny $T$}}$ operator all quantities we are concerned
about can be uniformly expressed in the same form except for
different positive conditions.

Consider the input state $\rho_0=|0\rangle\langle 0|^{\otimes 2}$.
We assume the POVM is  covariant and symmetric. Mathematically
this is equivalent to say that
\begin{eqnarray}
\mbox{Tr}(\hat{a}_{\phi}|0\rangle \langle 0|^{\otimes
2})=\mbox{Tr}(\hat{a}_0 |\phi\rangle\langle \phi|^{\otimes 2})
\,\,\, ( u(\phi)|0\rangle=|\phi\rangle)
\end{eqnarray}
and $\hat{a}_{\phi}$ is invariant under the permutation group
$S_2$. The covariance of the POVM can greatly simplify the
explicit form of $\hat{a}_0$.

Since the state $\rho_0$ is invariant under $u_g \otimes u_g$ with
\cite{ariano1}
\begin{eqnarray}
u_g= \left (
\begin{array}{c|c}
e^{i \theta_g} & \begin{array}{cc} 0 & \ldots \end{array} \\
\hline \begin{array}{c} 0 \\ \vdots \end{array} & u_g^{d-1}
\end{array}
\right ),
\end{eqnarray}
where $u_g^{d-1}$ is a $(d-1)$-dimensional unitary matrix with
$\mbox{det}(u_g^{d-1})=e^{-i \theta_g}$, the operator $\hat{a}_0$
must be commutable with $u_g \otimes u_g$. A detailed analysis can give the explicit
 form of $\hat{a}_0$. Here we choose $\hat{a}_0$ with the following form which is enough for our consideration
\begin{eqnarray}
\hat{a}_0 & = &I + \alpha (T^{(3)} \otimes I + I \otimes T^{(3)})
+ \beta T^{(3)} \otimes T^{(3)} \nonumber \\
&\mbox{}& + \gamma \sum_{m=1}^{d-1} (T_{0m}^{(1)} \otimes
T_{0m}^{(1)} + T_{0m}^{(2)} \otimes
T_{0m}^{(2)}) \nonumber\\
&   & \mbox{} + \delta \sum_{m,n=1}^{d-1} \Bigl ( T_{mn}^{(1)}
\otimes T_{mn}^{(1)} + T_{mn}^{(2)} \otimes T_{mn}^{(2)} \nonumber
\\
& & \mbox{} \>\>\>\>\>\>\>\>\>\>\>\>\>\>\>\>\>\>\>\>\>\>\>\> +
\frac{2}{d-2} T_{mn}^{(3)} \otimes T_{mn}^{(3)} \Bigr ),
\label{a0}
\end{eqnarray}
where $T^{(3)}=diag(1-d,\underbrace{1,1,\ldots,1}_{d-1})$,
$T_{mn}^{(1)}=|m \rangle \langle n| + |n \rangle \langle m|$,
$T_{mn}^{(2)}=-i |m \rangle \langle n| + i |n \rangle \langle m|$,
and $T_{mn}^{(3)}=|m\rangle \langle m| - |n\rangle \langle n|$.
Actually the last term of Eq. ($\ref{a0}$) corresponds to the
quadratic Casimir operator of unitary group $SU(d-1)$.

Since we have assumed that the POVM is covariant and symmetric,
the completeness relation can be reformulated as
\begin{eqnarray}
\int d u \, u^{\otimes 2}\hat{a}_0(u^\dag)^{\otimes2}=I,
\end{eqnarray}
where $du$ denotes the integration with respect to the normalized
Haar measurement \cite{book} of the unitary group. Taking into
account of the explicit form of $\hat{a}_0$, we can obtain
\begin{eqnarray}
\beta d(d-1)+4 \gamma(d-1)+2\delta d(d-2)=0. \label{norm}
\end{eqnarray}
 When we get a measurement
result $r$ corresponding to the operation $\hat{a}_r$, we can
guess the input state to be $|\phi_r\rangle$. The whole
information distilled from the measurement results can be
described by the following mean fidelity
\begin{eqnarray}
F & = & \int\! d\,\phi \sum_r \mbox{Tr}[\hat{a}_r \rho(\phi,\phi)]
| \langle \phi_r|\phi\rangle|^2 \nonumber \\
&=& \mbox{Tr}\left [
\hat{a}_0 \int d\,u \; u^{\otimes 2} \rho_0 (u^\dag)^{\otimes 2} |
\langle 0|u|0\rangle|^2 \right ]. \label{fidelity2}
\end{eqnarray}
Substituting Eq. ($\ref{a0}$) into Eq. ($\ref{fidelity2}$) we can
simplify the expression as
\begin{eqnarray}
F= \frac{1}{d}- \frac{ \alpha  (d+2)(d-1) - 2 \beta (d-1)(d-2)+
 \delta  d(d-2)} {2 d(d+1)(d+2)}. \nonumber \\ \label{fidelity3}
\end{eqnarray}

Now our aim is just to maximize the fidelity $F$ under the
constraints of Eq.($\ref{a0},\ref{norm}$) with $\hat{a}_0$
satisfying different positive conditions.

\emph{Case one}: When the input state is
$|\phi_{in}\rangle=|\phi\rangle|\phi\rangle$, positivity of
$\hat{a}_0$ gives
\begin{eqnarray}
\left \{
\begin{array}{l}
1-2 \alpha (d-1)+\beta(d-1)^2 \geq 0,  \\
1+2 \alpha + \beta + 2 \delta \frac{d-2}{d-1} \geq 0,  \\
1+2 \alpha + \beta -d(\alpha+\beta) \geq |2\gamma|, \\
1+2 \alpha + \beta -2\delta \frac{1}{d-1} \geq |2\delta|.
\end{array}
\right .
\end{eqnarray}
Maximizing $F$ now becomes a usual linear programming problem. A
simple algebra reveals the maximum of the fidelity
$F_{\parallel}=\frac{3}{d+2}$ is obtained at
$\alpha=-\frac{3}{4}$, $\beta=\frac{1}{2}$, $\delta=0$, and
$\gamma=-\frac{d}{4}$, which is consistent with the result of
Ref. \cite{brub}. The corresponding $\hat{a}_0$ now can be written
as
\begin{eqnarray}
\hat{a}_0 = \frac{d(d+1)}{2}|00\rangle \langle 00| + \frac{d}{2}
\sum_{i=1}^{d-1} |\psi_i\rangle \langle \psi_i|
\end{eqnarray}
with $|\psi_i\rangle=\frac{1}{\sqrt{2}}(|0i\rangle-|i0\rangle)$.


\emph{Case two}: For input state
$|\phi_{in}\rangle=|\phi\rangle|\phi^*\rangle$, the operator
$\hat{a}_0^{\mbox{\tiny $\sim$}\mbox{\tiny $T$}}$ should be
positive. This is equivalent to
\begin{eqnarray}
\left \{
\begin{array}{l}
1- \alpha (d-2)-\beta(d-1) \geq 0,  \\
1+2 \alpha + \beta - 2 \delta \frac{1}{d-1} \geq 0,  \\
1+2 \alpha + \beta + 2\delta \frac{d-2}{d-1} \geq |2\delta|, \\
\left ( 1+2 \alpha + \beta + 2\delta \frac{d-2}{d-1}\right ) \Bigl
[ 1-2 \alpha (d-1) + \beta(d-1)^2 \Bigr ] \geq |2\gamma|^2.
\end{array}
\right.  \label{case2}
\end{eqnarray}
Since Eq. ($\ref{case2}$) contains a nonlinear term, maximizing
$F$ corresponds to a nonlinear programming problem, which make the
question a little complicated. Before giving an analytical
expression, here we concentrate on a very special case. By setting
Eq. ($\ref{case2}$) to be equality constraints, one can find a
local extremal point of Eq. ($\ref{case2}$) is $
\alpha=\frac{1}{{\left( 1 + {\sqrt{1 + d}} \: \right) }^2}-1$, $
\beta=\frac{4 + \left(  d -2 \right)\! \left( d^2 + 2\,{\sqrt{1 +
d}} \, \right) }{d ^{2}  \left(   d -1 \right) }$, $
\delta=\frac{{\left(  {\sqrt{1 + d} -1} \right) }^2}{2\,d}$, and $
\gamma = \frac{2-d^2+(d-2)\sqrt{1+d}}{2d}$. This indicates that
$\hat{a}_0^{\mbox{\tiny $\sim$}\mbox{\tiny
$T$}}=|\psi_{local}\rangle \langle \psi_{local}|$ is a rank-1
operator with
\begin{eqnarray}
|\psi_{local} \rangle&=&\frac{1}{\sqrt{d}}\{
[(d-1)\sqrt{1+d}+1]|00\rangle \nonumber \\&& \mbox{}-
(\sqrt{1+d}-1)\sum_{i=1}^{d-1}|ii\rangle \}.
\end{eqnarray} The corresponding fidelity becomes
\begin{eqnarray}
F_{local}=\frac{2\,\left( 1 + 2\,d \right) }{\left( 1 + d \right)
\,\left( 2 + d \right) } -  \frac{\left(   d -1 \right) \,{\left(
{\sqrt{1 + d}} -1 \right) }^2}{d^2\,\left( 1 + d \right)  },
\end{eqnarray}
which is larger than $\frac{3}{d+2}$ for arbitrary integer $d \geq
2$. Interestingly, in the case of $d=2$, such rank-1 operator is
also global optimal \cite{massar}. Hence by considering the local
extreme value of $F$, it is enough to show that phase-conjugate
pair can encode more information than parallel pair.

In the general case, the global maximum of the fidelity for this
phase-conjugate input pair can be obtained when $\alpha =
\frac{A_--4}{4(d-1)}$, $\beta =\frac{(d-1)A_++4}{4(d-1)^2}$, $
\delta = \frac{A_-}{8(d-1)}$, and $ \gamma = -\frac{d}{4}$, with
$A_{\pm}=2d\pm\sqrt{2d(d+1)}$. The maximum fidelity can be
formulated as
\begin{eqnarray}
F_{\perp} = \frac{1}{d+2}\left ( 2+\sqrt{\frac{2d}{d+1}} \right ).
\end{eqnarray}
And the measurement operator now becomes
\begin{eqnarray}
\hat{a}_0^{\mbox{\tiny $\sim$}\mbox{\tiny
$T$}}=\frac{d}{2A_+}\sum_{i,j=1}^{d-1}|ij\rangle \langle ij| +
|\psi_{\perp}\rangle \langle \psi_{\perp}|,
\end{eqnarray}
with $ |\psi_{\perp}\rangle = \sqrt{\frac{dA_+}{2}}|00\rangle -
\sqrt{\frac{d}{2A_+}}\sum_{i=1}^{d-1}|ii\rangle$. In table. 1 we
give some explicit numerical results on these different
fidelities. One can easily get that $F_{\perp} \ge
F_{local}>F_{\parallel}$ and in the most case the local extrema
are very close to the corresponding global maxima.

\begin{table}\label{table-I}
\caption{Some numerical results of the information distilled from
two different input states (from Eq. ($\ref{case2}$)). Here $d$ is
the dimension of $\cal{H}$. Compared with parallel pair
($F_{\parallel}$), phase-conjugate input states ($F_{\perp}$ and
$F_{local}$) can encode more information for our figure of merits.
One can also find that the local extreme values are very close to
the global maxima.}
\begin{center}
\begin{tabular}{c|ccccccc}
  \toprule
  $d$ & 2 & 3 & 4 & 5 & 6 & 11 & 17  \\
  \colrule
  $F_{\parallel}$ & $0.75$ & $0.6$ & $0.5$ & $.4286$ &
  $.375$
  & $.2308$ & $.1579\phantom{\Big|}$ \\
 \colrule
 $F_{local}$ & $.7887$ & $.6444$ & $.5427$ & $.4678$ & $.4195$ & $.2531$ & $.1723$ \\
 \colrule
  $F_{\perp}$ & $.7887$ & $.6449$ & $.5442$ &
  $.4701$
  & $.4137$ & $.2580$ & $.1776\phantom{\Big|}$\\
  \botrule
\end{tabular}
\end{center}
\end{table}

It has been found that the optimal procedure to encode a quantum
state depends only on the dimension of the encoding space
\cite{bagan}. Phase-conjugate pair span the whole Hilbert space of
two particles while the space spanned by parallel pair is only
$\frac{d(d+1)}{2}$-dimension. So it might not be a surprise that
phase-conjugate pair encode more information. However, since the
optimal encoding state is often an entangled state, while in our
consideration, the two kinds of inputs are both direct-product
states, which make the whole question not so obvious.

Many interesting problems arise from these results. One may
consider, for example, the cloning machine with phase-conjugate
pair. It has been demonstrated that for a $2$-level system,
quantum cloning with anti-parallel spins can get better results
than parallel input spins when the number of the output copies is
sufficiently large \cite{anti-cloning}. Generally state estimation
can be considered as the limit case of quantum cloning
\cite{Asymptotic}, so it would be expected that in high dimension
case, quantum cloning with phase-conjugate pair can also get
higher fidelity. One can also consider generalizing this to the
case of having $N$ parallel input states and $M$ phase-conjugate
states \cite{relative}. Another point is that we have only
considered the case that the input states are $n$-fold, actually
to find the condition under which quantum operation is bounded by
classical information is still an interesting question.

In conclusion, we have shown that an antilinear map is essentially
a classical operation for $N$ parallel input states. The fidelity
of such operation is bounded by the classical information
distillable from the input state. We have also considered the
classical information contained in two different input states ($|
\phi \rangle |\phi \rangle$ and $|\phi \rangle | \phi^* \rangle$).
Compared with the parallel pair, more information can be encoded
in phase-conjugate pair for our figure of merit. We expect our
work will be helpful to explore the role played by antilinear map
within quantum information.

This work was funded by the National Fundamental Research Program
(2001CB309300), the National Natural Science Foundation of China
(10304017, 60121503), the Innovation Funds from the Chinese
Academy of Sciences, and Program for New Century Excellent Talents
in University.

Note added: Very recently J. Fiur\'{a}\v{s}ek \cite{fij} has got
similar results about the state estimation from phase-conjugate
pair $|\phi\rangle |\phi^*\rangle$, and he has also given a
physical explanation about the results that we have got in this
work. Interestingly the global optimal fidelity we got corresponds
to the result of the optimal probabilistic estimation strategy,
while the local extreme value agrees with the result of the
optimal deterministic estimation strategy.

\end{document}